\documentclass[twocolumn,showpacs,preprintnumbers,amsmath,amssymb,prl,superscriptaddress]{revtex4}

\usepackage{graphicx}
\usepackage{dcolumn}
\usepackage{bm}

\begin{document}

\preprint{APS/123-QED}

\title{X-ray Magnetic Circular Dichroism of Valence Fluctuating State in Eu \\
at High Magnetic Fields
}

\author{Y.~H.~Matsuda}
 \altaffiliation[Also at ]{PRESTO, Japan Science and Technology Agency, Saitama 332-0012}
\altaffiliation[Present address, ]{Institute for Solid State Physics, University of Tokyo, Chiba 277-8581}
\email{ymatsuda@issp.u-tokyo.ac.jp}
\author{Z.~W.~Ouyang}
\author{H.~Nojiri}
\affiliation{%
Institute for Materials Research, Tohoku University, 2-1-1 Katahira, Aoba-ku, Sendai 980-8577, Japan
}%

\author{T.~Inami}
\author{K.~Ohwada}
\affiliation{
Synchrotron Radiation Research Unit, Japan Atomic Energy Agency, Sayo, Hyogo 679-5148, Japan 
}%
\author{M.~Suzuki}
\author{N.~Kawamura}
\affiliation{SPring-8/JASRI,1-1-1 Kouto, Sayo, Hyogo 679-5198 Japan 
}%

\author{A.~Mitsuda}
\author{H.~Wada}
\affiliation{
Department of Physics, Kyushu University, Fukuoka 812-8581, Japan 
}%

\date{\today}

\begin{abstract}
X-ray magnetic circular dichroism (XMCD) at the Eu $L$-edge (2$p$ $\rightarrow$ 5$d$) in 
two compounds exhibiting valence fluctuation, namely EuNi$_2$(Si$_{0.18}$Ge$_{0.82}$)$_2$ and EuNi$_2$P$_2$, has been investigated 
at pulsed high magnetic fields of up to 40~T. 
A distinct XMCD peak corresponding to the trivalent state (Eu$^{3+}$; $f^6$), whose ground state is nonmagnetic ($J$=0), was observed 
in addition to the main XMCD peak corresponding to the magnetic ($J$=7/2) divalent state (Eu$^{2+}$; $f^7$).
This result indicates  
that the $5d$ electrons belonging to both valence states are magnetically polarized. 
It was also found that the ratio $P_{5d}$(3+)/$P_{5d}$(2+) between the polarization of 5$d$ electrons ($P_{5d}$) in the Eu$^{3+}$ state and that of Eu$^{2+}$ 
is $\sim$0.1 in EuNi$_2$(Si$_{0.18}$Ge$_{0.82}$)$_2$ and $\sim$0.3 in EuNi$_2$P$_2$ at magnetic fields where 
their macroscopic magnetization values are the same. 
The possible origin of the XMCD of the Eu$^{3+}$ state and an explanation of the dependence of 
$P_{5d}(3+)$/$P_{5d}(2+)$ on the material are discussed in terms of hybridization between the conduction electrons and the 
$f$ electrons.

\end{abstract}

\maketitle
At low temperatures, where thermal excitations are suppressed, quantum effects become clear 
and intriguing phenomena such as quantum phase transitions are observed 
in high magnetic fields \cite{Harrison03, Millis02}.
However, 
microscopic measurement techniques are limited in the presence of high magnetic fields. For example, photoemission experiments 
at high magnetic fields are very difficult to perform.
A synchrotron X-ray is an element- and shell-selective microscopic probe, and X-ray magneto-spectroscopy is a powerful 
method for studying electronic states at high magnetic fields.
Recently, X-ray absorption spectroscopy (XAS) at high magnetic fields up to 40~T has been carried out 
in order to clarify field-induced valence transitions of 
Yb- and Eu-based intermetallic compounds by using a pulsed magnet \cite{Matsuda07, Matsuda08}. 
%
As for the research of magnetic materials, X-ray magnetic circular dichroism (XMCD) spectroscopy is 
an even more powerful technique than XAS.
Since high-magnetic-field XMCD can be used for examining antiferromagnetic and paramagnetic materials 
\cite{Mathon07} as well as ferromagnetic materials,  
various kinds of field-induced phenomena, such as metamagnetic transitions, can be studied microscopically.

Valence fluctuation phenomena found in rare-earth intermetallic compounds 
have been attracting considerable attention from the viewpoint of the strong correlation 
of electrons.
In the valence fluctuation state at low temperatures, the Kondo effects becomes significant and 
the ground state is nonmagnetic in many cases.
When magnetic fields are applied, the localized magnetic moment tends to appear.
In fact, field-induced valence transitions are found along with the metamagnetism in some systems exhibiting valence fluctuation, such as 
Ce$_{0.8}$La$_{0.1}$Th$_{0.1}$ \cite{Drymiotis05}, YbInCu$_4$\cite{Yoshimura88}, 
EuNi$_2$(Si$_{1-x}$Ge$_x$)$_2$\cite{Wada97}.
In order to elucidate the valence fluctuation phenomena, the XMCD experiment is highly intriguing.
We can examine the magnetic polarization of electrons in different valence states determined by the magnetic fields.

In this Letter, we present the application of high-magnetic-field XMCD to EuNi$_2$(Si$_{0.18}$Ge$_{0.82}$)$_2$ and 
EuNi$_2$P$_2$, which are typical materials exhibiting valence fluctuation \cite{Wortmann91, Nagarajan85}.
The magnetic polarization of the Eu 5$d$ electrons in different valence states is examined by using XMCD spectroscopy 
at high magnetic fields. 
The dependence of the polarization of the 5$d$ electrons on the field and the material is discussed in terms of  
hybridization between the conduction and $f$ electrons ($c$-$f$ hybridization).

%
The XMCD experiment with Eu $L_{2,3}$-edge (2$p_{1/2,3/2}\rightarrow 5d$) was carried out 
at high magnetic fields at BL39XU in SPring-8\cite{Maruyama99} by using a miniature pulsed magnet 
producing fields of up to 40~T \cite{Matsuda07, Matsuda08}.
A polycrystal of EuNi$_2$(Si$_{0.18}$Ge$_{0.82}$)$_2$ 
and a single crystal of EuNi$_2$P$_2$ 
were used.
The crystals were powdered and diluted in order to achieve an effective 
sample thickness of about 10~$\mu$m for the transmission measurement.
%
%
%
An experimental setup for high-field XMCD spectroscopy was similar to that for high-field XAS, described in 
Ref.~\cite{Matsuda07,Matsuda08}.  
A diamond X-ray phase plate was introduced into the present setup to generate circularly polarized X-rays.  
A storage oscilloscope was used to record the detector output voltages corresponding to the incident 
and transmitted X-ray intensities of the sample as a function of time, together with the pulsed magnetic field.  
The XMCD signal ($\Delta \mu t =\mu^+ t - \mu^- t$) is determined as the difference in the absorption intensities 
for right- ($\mu^+ t $) and left- ($\mu^- t $) circular polarization. The $\mu^+ t $ and $\mu^- t $ were measured 
by successive two shots of a pulsed field at a fixed X-ray energy, as the photon helicity was reversed every shot.  

\begin{figure}
\includegraphics[width=6cm]{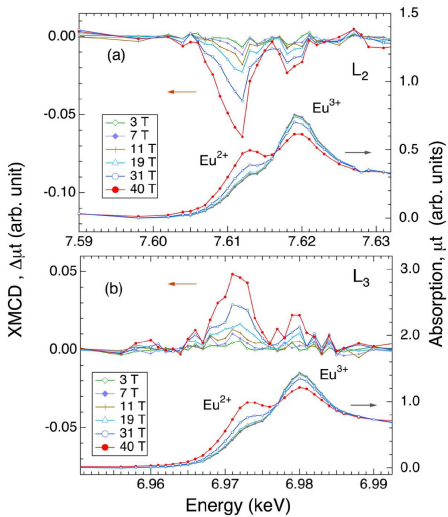}
\caption{\label{fig:mcdENSG} (a) XMCD and XAS spectra at various magnetic fields at 5~K for the $L_2$ edge. 
XAS spectra are plotted in such a way that the absorption at 7.598 keV becomes zero. 
(b) XMCD and XAS spectra at various magnetic fields at 5~K for the $L_3$ edge. XAS spectra are plotted in 
such a way that the absorption at 6.956 keV becomes zero.}
\end{figure}

%
The measured XMCD spectra of EuNi$_2$(Si$_{0.18}$Ge$_{0.82}$)$_2$ are shown in Fig.~\ref{fig:mcdENSG} 
together with the XAS spectra at several magnetic fields.
The XAS spectra of the $L_2$ and $L_3$ absorption edges are very similar, except for the intensities.
The valence fluctuation manifests itself as a double-peak structure in the XAS spectra \cite{Wortmann89}. 
The higher energy XAS peak, which is shown in Fig.~\ref{fig:mcdENSG}, is attributed to the Eu$^{3+}$ state, 
while the lower energy peak is attributed to the Eu$^{2+}$ state \cite{Wortmann91, Wada97}. 
The XAS spectrum strongly depends on the magnetic field, exhibiting a valence change induced by 
the magnetic field 
\cite{Matsuda08, Scherzberg84}.
We found that the double-peak structure corresponding to the two valence 
states is observed in 
the XMCD spectra as well as in the XAS spectra.
The double peaks indicate that the electrons in the Eu 5$d$ orbital, 
where the excited electron finally goes in the $L_{2,3}$ transitions, 
are magnetically polarized not only in the Eu$^{2+}$  state ($f^7$; $J=7/2$ in the ground state), but also in 
the Eu$^{3+}$  state ($f^6$; $J=0$ in the ground state).

\begin{figure}
\includegraphics[width=6cm]{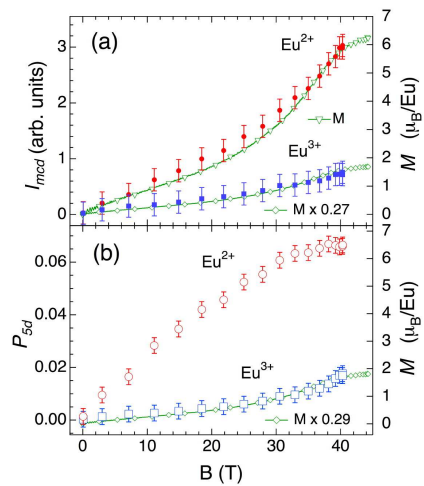}
\caption{\label{fig:Hdep} (a) Magnetic field dependence of the integrated XMCD intensities of Eu$^{2+}$ and 
Eu$^{3+}$ states for the $L_2$ edge. (b) The degree of magnetic polarization of 5$d$ electrons 
is plotted as a function of the magnetic field.
Magnetization measured at 4.2~K \cite{MitsudaPhD} is shown for comparison.}
\end{figure}

Figure ~\ref{fig:Hdep} shows the dependence of the 
integrated intensity of the XMCD peak ($I_{mcd}$) on the magnetic field, as well as the magnetic 
polarization of the 5$d$ electrons ($P_{5d}$). 
The degree of magnetic polarization of the Eu 5$d$ electrons in each valence state is defined in this paper as 
$P_{5d}=\int\Delta \mu t \; dE /\int \mu t\; dE$, where $\Delta \mu t$  and $\mu t= (\mu^+ t + \mu^- t)/2$ 
are the XMCD intensity and the absorption intensity, respectively. $\int\Delta \mu t \; dE$ (=$I_{mcd}$) is 
deduced by integration of the XMCD peak shown in Fig.~\ref{fig:mcdENSG} for each valence state.
$\int \mu t \; dE $ is obtained from the integrated intensity of the absorption peak. 
Curve fitting analysis is performed in order to evaluate the absorption intensities $\int \mu t \; dE =I_2$ and $I_3$ for the 
Eu$^{2+}$ and Eu$^{3+}$ states, respectively, at different magnetic fields. 
The Eu valence $v^*$ is directly deduced from $v^*$=2{$I_2$/($I_2$+$I_3$)}+3{$I_3$/($I_2$+$I_3$)} \cite{Wortmann89}. 
The details of the curve fitting are shown in Ref. 4. 
The $v^*$ values at 5~K obtained in the present study namely 2.74 at 0~T and 2.47 at 40~T are in good agreement with 
the field dependence of $v^*$ reported in our previous paper \cite{Matsuda08}. 
%
%
%
%
The magnetization ($M$) of the sample at 4.2~K \cite{MitsudaPhD} is shown in Fig.~\ref{fig:Hdep} in order 
to compare it with the dependence of $I_{mcd}$ and $P_{5d}$ on the field.
An increase of the magnetization caused by the valence change is observed 
at around 35~T \cite{MitsudaPhD, Wada97}. 
In addition, $I_{mcd}$ of Eu$^{2+}$ and that of Eu$^{3+}$ scale together with the magnetization, 
as expected for XMCD experiments.
However, $P_{5d}$ is qualitatively different for the Eu$^{2+}$ and Eu$^{3+}$ states.
If we define the $P_{5d}$ of Eu$^{2+}$($f^7$) and that of Eu$^{3+}$($f^6$) as $P_{5d}$(2+) and 
$P_{5d}$(3+), respectively, it is found that the dependence of $P_{5d}$(2+) on the field is convex upward 
and exhibits saturation more clearly than the magnetization at around 40~T, while $P_{5d}$(3+) still 
appears to follow the magnetization curve. 
Since the magnetic field dependence of $I_{mcd}$ and $P_{5d}$ for the $L_3$-edge is found to be 
qualitatively similar to that for the $L_2$-edge, we only show and discuss the results of the $L_2$-edge 
in this Letter.

\begin{figure}
\includegraphics[width=6cm]{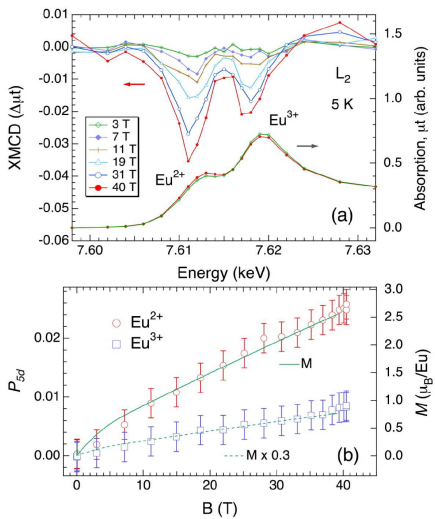}
\caption{\label{fig:mcdENP} (a)XMCD spectra in EuNi$_2$P$_2$ at high magnetic fields. 
The XAS spectra at 3 and 40~T, which are plotted in such a way that the absorption intensity 
becomes zero at 7.598 eV, are also shown. (b) The degree of magnetic polarization of the 5$d$ electrons, 
as well as the magnetization measured at 4.2~K, are plotted as functions of the magnetic field.}
\end{figure}

%
In Fig.~\ref{fig:mcdENP}(a), we show the results for another sample, namely EuNi$_2$P$_2$.
The Eu valence in EuNi$_2$P$_2$ is $2.5 - 2.6$ \cite{Perscheid85, Nagarajan85}, which is almost exactly midway between 
2 and 3, due to the strong hybridization, and this material is considered to be characterized by stronger hybridization than 
EuNi$_2$(Si$_{0.18}$Ge$_{0.82}$)$_2$.
We found two clear peaks in the XAS and XMCD spectra, as in the case of EuNi$_2$(Si$_{0.18}$Ge$_{0.82}$)$_2$.
Only small changes in the XAS spectra are visible even at 40~T, and the Eu valence depends only slightly on the magnetic field in 
the case of EuNi$_2$P$_2$ (2.64 at 0~T and 2.61 at 40~T). 
The deduced polarization of the 5$d$ electrons ($P_{5d}$) for each valence state is plotted as a function of the field in 
Fig.~\ref{fig:mcdENP} (b) together with the magnetization. 
It is found that both $P_{5d}$ values follow the magnetization.

%
In the following paragraphs, we discuss the XMCD spectra and the deduced polarization $P_{5d}$. 
The double-peak structure found in the XMCD spectra of EuNi$_2$(Si$_{0.18}$Ge$_{0.82}$)$_2$ and EuNi$_2$P$_2$ 
indicates that the Eu 5$d$ electrons are magnetically polarized not only in the Eu$^{2+}$ state, but also in the 
Eu$^{3+}$ state.
Since the ground state of Eu$^{2+}$ ($f^7$) is magnetic ($J$=7/2; $L=0$, $S=7/2$), we 
expect significant XMCD for the Eu$^{2+}$ state.
However, the ground state of Eu$^{3+}$ ($f^6$) is nonmagnetic ($J$=0; $L=3$, $S=3$) \cite{Wada97, Wortmann91}, 
and it is not obvious why there is a finite XMCD signal for the Eu$^{3+}$ state.
Possible origins for the XMCD signal of the Eu$^{3+}$ state include (i) hybridization of the Eu$^{2+}$ and Eu$^{3+}$ states in the 
final state, 
 (ii) effect of the excited state of Eu$^{3+}$ state ($J=1$), 
 and (iii) magnetic polarization of the conduction electrons.

Regarding the first option (i), we can reasonably assume that there is hardly any mixing of the valence in the final state of 
the $L_{2,3}$ transitions. 
This is due to the fact that the energy difference between the 5$d$ electrons in Eu$^{3+}$ and Eu$^{2+}$ is around 8~eV 
owing to the strong Coulomb interaction between the core-hole and the $f^n$ electrons ($n$=6 or 7).
Therefore, it can be assumed that each absorption peak can be attributed to a nearly pure Eu$^{3+}$ or Eu$^{2+}$ state.
In this case, the possibility of obtaining XMCD due to the Eu$^{2+}$ component from the absorption assigned to the Eu$^{3+}$ 
state is almost zero. 

Since the first excited state of Eu$^{3+}$ is magnetic ($J$=1)
\cite{Wada97, Wortmann91}, the second possible origin 
should have some effects. 
The energy separation between the first excited and ground states is about 480 K \cite{Wada97}.
When we estimate the magnetization due to the Van Vleck paramagetism using  
$1.43\times 10^{-6}$ ($\mu_B$/Eu$^{3+}$)/T as the magnetic susceptibility \cite{Frank32}, 
the magnetization of Eu$^{3+}$ is estimated to be about 0.6 $\mu _B$/ Eu$^{3+}$ at 40~T.
This value is in good agreement with the Van Vleck term reported in a previous work on 
Eu$_3$Fe$_5$O$_{12}$ \cite{Mizumaki05}.
%
If we take the Van Vleck term of Eu$^{3+}$ into account, the relative contribution of the magnetization 
between the two valence states at fields lower than 20~T (below the transition field of 
EuNi$_2$(Si$_{0.18}$Ge$_{0.82}$)$_2$), $M$(3+)/$M$(2+) is assumed to be around 0.09 for 
EuNi$_2$P$_2$ and 0.06 for  EuNi$_2$(Si$_{0.18}$Ge$_{0.82}$)$_2$.
%
It appears that these values are two or three times smaller than the ratio $P_{5d}$(3+)/$P_{5d}$(2+), as we show below.
%

However, in compounds exhibiting valence fluctuation, the mixing between the Eu$^{3+}: 
J=0 $ and Eu$^{3+}: J=1$ states can be enhanced by the strong  $c$-$f$ hybridization.
%
According to a previous work on EuCu$_2$Si$_2$\cite{Rohler82}, significant contribution of the Eu$^{3+}: J=1$ state 
to the magnetic properties was observed.
Therefore, the effect of Eu$^{3+}: J=1$ state is one of the plausible origins of the Eu$^{3+}-$XMCD observed 
in the present work.

The third possible origin is that the conduction electrons are polarized by strong hybridization with the $4f$ electrons 
in the magnetic Eu$^{2+}$ states, and these polarized conduction electrons induce the magnetic polarization of the 
5$d$ electrons in the nonmagnetic Eu$^{3+}$ state.
Owing to the strong $c$-$f$ hybridization, this mechanism also can be regarded as plausible.
Moreover, this postulated mechanism is similar to that found in DyLu, where magnetic polarization of the 5$d$ 
electrons in nonmagnetic Lu ($f^{14}$) in DyLu was observed by magnetic resonance scattering \cite{Everitt95}. 
%
Here, it should be stressed that the strength of the $c$-$f$ hybridization plays an important role for the 
XMCD of the Eu$^{3+}$ state not only in the second possible origin but also in the third possible origin.

Next, we compare the relative magnitude $P_{5d}$(3+)/$P_{5d}$(2+) of the polarization of 5$d$ electrons 
between EuNi$_2$P$_2$ and EuNi$_2$(Si$_{0.18}$Ge$_{0.82}$)$_2$ in the same magnetization range, 
$1<$ $M$/($\mu _B$/Eu) $<3$.
From Figs. \ref{fig:Hdep}(b) and  \ref{fig:mcdENP}(b), we obtain $P_{5d}$(3+)/ $P_{5d}$(2+)  =$ 0.30\pm 0.03$ 
and $ 0.12 \pm 0.03$ for EuNi$_2$P$_2$ and EuNi$_2$(Si$_{0.18}$Ge$_{0.82}$)$_2$, respectively.
Those values are nearly constant in the interval $1<$ $M$/($\mu _B$/Eu) $<3$. 
The ratio of the values for the two materials is 0.30/0.12=2.5.
This dependence on the material suggests that the induced polarization of the 5$d$ electrons in the Eu$^{3+}$ state 
depends on the electronic structure, as well as possibly on the strength of the $c$-$f$ hybridization.
%

If we assume a simple two-level scheme for describing the hybridization and denote the energy gap 
between the two valence states as $\Delta E$, the hybridization energy parameter $V$ can be evaluated 
from the Eu valence $v^*$ at low temperatures and zero magnetic fields.
If we use $v^*$=2.6 for EuNi$_2$P$_2$ and 2.8 for EuNi$_2$(Si$_{0.18}$Ge$_{0.82}$)$_2$, 
it is found that  $V/\Delta E$=2.3 and 0.7 for EuNi$_2$P$_2$ and EuNi$_2$(Si$_{0.18}$Ge$_{0.82}$)$_2$,  
respectively.
The relative value of the hybridization strength between the two materials is thus estimated to be 2.3/0.7, 
which is around 3.3.
Therefore, we found that the relative value of $P_{5d}$(3+)/$P_{5d}$(2+) in the two materials 
(2.5) is close to the estimated relative hybridization value in the two materials (3.3).
This agreement might support our assumption that Eu$^{3+}$-XMCD is induced through the $c$-$f$ hybridization.

Finally, we discuss the dependence of $P_{5d}$ in EuNi$_2$(Si$_{0.18}$Ge$_{0.82}$)$_2$ on the magnetic field.
Since the field dependence of $P_{5d}$(3+) seems to follows the macroscopic magnetization curve, 
as seen in Fig.~\ref{fig:Hdep}, it appears that 
$P_{5d}$(3+) reflects the magnetization contributed from many Eu sites through the $c$-$f$ hybridization.
Hence, it is likely that the magnetic moments due to the $J$=7/2 (Eu$^{2+}$) and $J$=1 (Eu$^{3+}$) states of 
neighboring Eu sites induce the polarization of 5$d$ electrons of Eu$^{3+}$ state.
%
%
Regarding the Eu$^{2+}$ state, since the 5$d$ electrons are magnetically polarized by the local $d-f$ 
exchange interaction, $P_{5d}$(2+) directly reflects the magnetic polarization of the 4$f$ electrons in 
the Eu$^{2+}$ state.
Hence, we regard the dependence of $P_{5d}$(2+) on the magnetic field as a hypothetical magnetization 
curve when all Eu sites have localized $J$=7/2 magnetic moments. 
It is noteworthy that the magnetization curve of antiferromagnetic EuNi$_2$(Si$_{0.05}$Ge$_{0.95}$)$_2$ 
\cite{Wada97} is similar to that of the dependence of $P_{5d}$(2+) on the magnetic field.


%

In conclusion, valence-selective XMCD has been clearly observed in Eu-based compounds exhibiting valence fluctuation, namely 
EuNi$_2$(Si$_{0.18}$Ge$_{0.82}$)$_2$ and EuNi$_2$P$_2$, for fields of up to 40~T.
Finite XMCD is observed for both valence states Eu$^{2+}$ and Eu$^{3+}$, although the expected magnetic properties 
are very different, i.e., the ground state of Eu$^{2+}$ has $J$=7/2, while that of Eu$^{3+}$ has $J$=0.
We have proposed possible explanations for the XMCD of the nonmagnetic Eu$^{3+}$ state in terms of 
hybridization between the conduction electrons and the 4$f$ electrons ($c$-$f$ hybridization)
; the two possible origins are 
(1) mixing of Eu$^{3+}$ $J=1$ state through the $c$-$f$ hybridization and (2) spin polarization of the conduction 
electrons due to the hybridization effect.
%
%
%
%
%
Another intriguing finding concerns the fact that the dependence of the magnetic polarization of the 
5$d$ electrons ($P_{5d}$) on the magnetic field 
is qualitatively different for the Eu$^{2+}$ and Eu$^{3+}$ states at magnetic fields 
higher than the valence transition field of EuNi$_2$(Si$_{0.18}$Ge$_{0.82}$).
This difference is attributed to the different origins of the XMCD, where the Eu$^{2+}$ XMCD is caused by localized 
electrons, while Eu$^{3+}$ XMCD reflects the character of the itinerant electrons.
Although the detailed mechanism of the XMCD in Eu-based materials exhibiting valence fluctuation is still unclear, it is 
plausible that the $c$-$f$ hybridization plays an important role.
In addition, actually, recent photoemission experiments reported evidence suggesting strong hybridization between 
Eu 4$f$ and Ni 3$d$ electrons in EuNi$_2$P$_2$ \cite{Danzenbacher09}. 
Hence, the conduction electrons referred to in the present XMCD paper can be Ni 3$d$ electrons.
For the better understanding of the phenomenon, a theoretical model beyond a single atomic picture \cite{Kotani07} 
might be required.
%
%
%
%
%

This work is partly supported by a Grant-in-Aid for Scientific Research on Priority 
Area "High Field Spin Science in 100 T" (No.451) provided by the Ministry of Education, Culture, Sports, Science and Technology (MEXT), Japan.
Y.H.M. and H.N. thank Prof. I.~Harada, Prof. A. Kotani and Prof. T.~Ziman for fruitful discussions.

\end{document}